\documentclass[10pt,conference,twocolumn,]{IEEEtran}
\IEEEoverridecommandlockouts

\usepackage{amsmath,amssymb,amsfonts}
\usepackage{cite}

\usepackage{graphicx}
\usepackage{textcomp}
\usepackage{xcolor}
\usepackage{times}
\usepackage{subfigure}
\usepackage{amssymb,amsmath}
\usepackage{acronym}  
\usepackage{balance}

\usepackage{bm}
\usepackage{algorithm}
\usepackage{algorithmic}

\usepackage{lipsum}
\usepackage{stfloats}

\allowdisplaybreaks[4]

\def\BibTeX{{\rm B\kern-.05em{\sc i\kern-.025em b}\kern-.08em
    T\kern-.1667em\lower.7ex\hbox{E}\kern-.125emX}}

\begin{document}
\title{Channel Feedback for Reconfigurable Intelligent Surface Assisted Wireless Communications}
\author{
	\vspace{0.2cm}
	\IEEEauthorblockN{
		Decai~Shen
		and
		Linglong~Dai}
	\IEEEauthorblockA{\IEEEauthorrefmark{0}
		Beijing National Research Center for Information Science and Technology (BNRist)\\
		Department of Electronic Engineering,
		Tsinghua University,
		Beijing 100084, China\\
		Emails: sdc18@mails.tsinghua.edu.cn, daill@tsinghua.edu.cn
	}
}
\maketitle
\begin{abstract}
Reconfigurable intelligent surface (RIS) has received widespread attention owing to the superiority of changing the wireless propagation environment intelligently. Channel feedback is essential in frequency division duplex (FDD) RIS-assisted wireless communications, since the downlink channel state information (CSI)  needs to be acquired by the base station (BS) from the user equipment (UE) for the joint beamforming  at the BS and the RIS. In this paper, we exploit the single-structured sparsity of  RIS channel, which means that the  sparse angular-domain channels  of different users share the same structured sparsity only in the column dimension, but not in the row dimension. Based on this  characteristic, we propose a dimension reduced channel feedback scheme with the low overhead. Specifically, the downlink CSI can be expressed with the structured information and unstructured information. The structured information of the multi-user channels can be fed back to the BS only by one user according to  the single-structured sparsity, then the unstructured information can be fed back with a fairly low overhead by different users, respectively. Moreover, by utilizing the angle coherence time, the per-user overhead can be reduced further, while the near-optimal performance can still be guaranteed.   Simulation results show that,  the proposed  scheme can reduce the channel feedback overhead by 80\% compared to the conventional method.


\end{abstract}
\section{Introduction}\label{S1}
Reconfigurable intelligent surface (RIS) has been recently recognized as a promising technique for future 6G communications~\cite{Ding6G}. Instead of adapting to the propagation environment in existing wireless communication systems, RIS can change the propagation environment by leveraging its controllable metamaterial-based elements. To jointly design the beamforming  at the base station (BS) and the RIS~\cite{Beamforming3,Bichai},  it is essential for the BS to acquire the downlink channel state information (CSI). In the frequency division duplex (FDD) aided wireless communications, we need to design a channel feedback scheme for user equipments (UEs) to feed the downlink CSI back to the BS with the channel codebook. Unfortunately, the size of the conventional codebook for channel feedback exponentially increases with the number of RIS elements and BS antennas~\cite{FBSubspace1(ShenTcom)}, which causes unbearable overhead for channel feedback in practical RIS-assisted wireless communication systems.

Although channel feedback has not been investigated for RIS-assisted wireless communications, it has been widely studied in current wireless communications~\cite{FBSubspace1(ShenTcom),FBCodebook1(RVQ)}. The downlink CSI should be accurately fed back to the BS, after channel estimation for  downlink CSI by the UE at first. To feed back the downlink CSI, an appropriate codeword is selected from a pre-set codebook at the UE to represent the downlink CSI as accurately as possible. Then, the index of this codeword is fed back to the BS via the uplink channel, which is usually assumed as a perfect known channel~\cite{FBCodebook1(RVQ)}. Based on the  pre-set codebook and codeword index, the  downlink CSI can be recovered at the BS. However, the existing channel feedback schemes cannot be directly applied to the RIS-assisted wireless communications, since the channel size is much larger (e.g., $64\times48$~\cite{ChongwenNumberOfRISElements} with single-antenna UE) than the current scenario without RIS (e.g., $64\times1$~\cite{FBSubspace1(ShenTcom)}), which will result in the pretty high feedback overhead and large codebook size. Besides, most of the existing codebooks  consist of one-dimension vector codewords, which are not applicable for two-dimension channel feedback in RIS-assisted wireless communications.

In this paper, channel feedback is investigated for the first time in RIS-assisted wireless communications  to the best of our knowledge. Particularly, we exploit the single-structured sparsity for  RIS-assisted wireless communication systems. The channel between the RIS and the UE is \emph{user-specific},  while the channel between the BS and the RIS is \emph{user-independent}. Based on this,  we can find that  different users' sparse angular-domain channels share the same structured sparsity only in the column dimension, but not in the row dimension. This characteristic is termed as the single-structured sparsity in this paper. Then, we propose a dimension reduced  channel feedback scheme with low overhead for the RIS-assisted wireless communications.  With the proposed scheme based on the  single-structured sparsity,  the downlink CSI can be expressed with the structured information
and unstructured information. The structured information of different users' channels can be fed back to the BS only by one user, then the unstructured  information  can be fed back with a  fairly low overhead by different users, respectively.   Simulation results verify that the proposed channel feedback scheme can  obviously reduce the per-user channel feedback overhead in  RIS-assisted wireless communication systems.


{\it Notations}: Lower-case and upper-case boldface letters stand for  vectors and  matrices, respectively; ${{{(\cdot)}^T}}$, ${{{(\cdot)}^H}}$, and ${{{(\cdot)}^{ - 1}}}$ denote the transpose, conjugate transpose, and inversion of a matrix, respectively;   $\sin ^{2}\left(\measuredangle\left(\mathbf{a}, \mathbf{b}\right)\right)=1-\frac{\left|\mathbf{a}^{\mathrm{H}} \mathbf{b}\right|^{2}}{\|\mathbf{a}\|^{2}\|\mathbf{b}\|^{2}}$; $\otimes$ denotes the kronecker product operator; $\mathbf{{A}}_{(:,n)}$ denotes the $n$-th column of the matrix $\mathbf{{A}}$. $\mathrm{E}\left[\cdot\right]$ denotes the expectation operator.

\begin{figure}[tp]
\begin{center}
\vspace*{3mm}\includegraphics[width=1\linewidth]{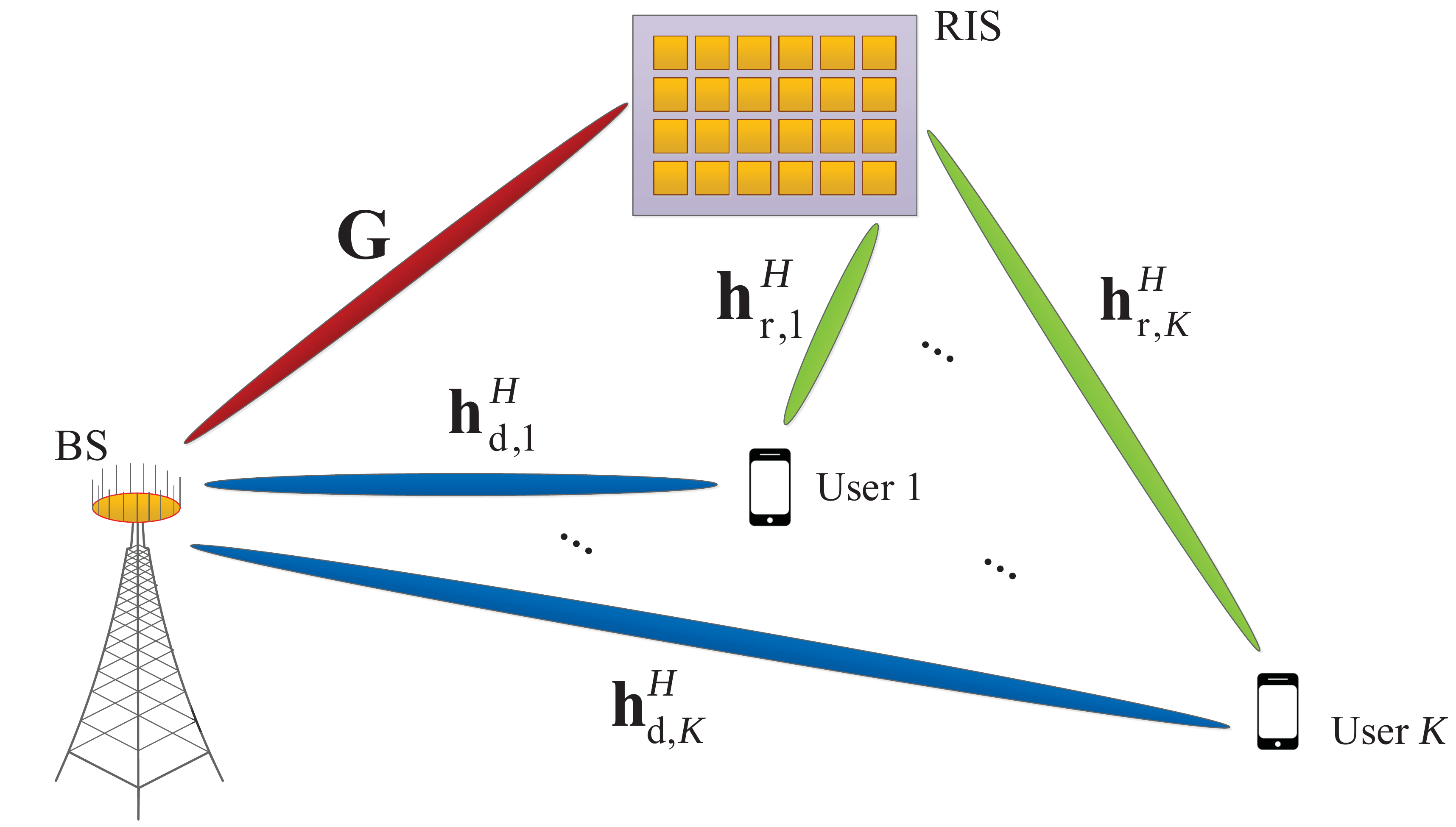}
\end{center}
\vspace*{-0mm}\caption{RIS-assisted wireless communication system.} \label{FIG1}
\vspace*{0mm}
\end{figure}


\section{System Model}\label{S2}


In this paper, a narrowband RIS-assisted wireless communication system is considered as illustrated in Fig. \ref{FIG1}, where  a BS with ${M}$ antennas is assisted by an RIS with ${N}$ elements to simultaneously serve ${K}$ single-antenna users. Hence, the downlink signal model for the $k$-th user can be expressed as~\cite{NadeemSystemModel}
\begin{equation}\label{eq17}
\begin{aligned}
y_{k} =\left({\mathbf{h}}^{H}_{d,k}+{\mathbf{h}}^{H}_{r,k}{\mathbf{\Phi}}{\mathbf{G}}\right) \mathbf{x}+n_{k},
\end{aligned}
\end{equation}
where $y_{k}$ is the received signal of the $k$-th user,  ${\mathbf{h}^H_{d,k}\in{\mathbb{C}}^{1 \times M}}$, ${\mathbf{h}^H_{r,k}\in{\mathbb{C}}^{1 \times N}}$, and ${\mathbf{G}\in{\mathbb{C}}^{N \times M}}$ denote the direct BS-UE channel from the BS to the ${k}$-th user, the RIS-UE channel from the RIS to the ${k}$-th user, and the BS-RIS channel matrix, respectively; $\mathbf{x}\in\mathbb{C}^{M\times1}$ is the precoded transmitted signal at the BS, and $n_{k} $ is the complex Gaussian noise at the $k$-th user with zero
mean and unit variance. Particularly, ${\mathbf{\Phi}}\in{\mathbb{C}}^{N \times N}$ represents the phase-shift diagonal matrix of the RIS as ${\mathbf{\Phi}}= {\mathrm{diag}}\left(\bm{\phi}^T\right)={{\mathrm{diag}}}\left({\phi}_1, \cdots, {\phi}_n, \cdots, {\phi}_N\right)$, where ${\phi}_n= e^{j\varphi_n}$ $\left(\varphi_n \in [0,2\pi], n=1,2,\cdots,N\right)$ represents the phase shift of the ${n}$-th RIS element.

By utilizing the property of diagonal matrix, i.e.,  ${\mathbf{h}}^{H}_{r,k}{\mathbf{\Phi}}{\mathbf{G}} ={\bm{\phi}^{T}}\mathrm{diag}\left({\mathbf{h}}^{H}_{r,k}\right){\mathbf{G}}$, the equivalent baseband downlink channel ${\mathbf{h}^H_{\mathrm{DL},k}\in{\mathbb{C}}^{1 \times M}}$ for the ${k}$-th user can be expressed as
\begin{equation}\label{eq3}
\begin{aligned}
{{\mathbf{h}}^{H}_{\mathrm{DL},k}}={\mathbf{h}}^{H}_{d,k}+{\mathbf{h}}^{H}_{r,k}{\mathbf{\Phi}}{\mathbf{G}} ={\mathbf{h}}^{H}_{d,k}+{\bm{\phi}^{T}}\mathrm{diag}\left({\mathbf{h}}^{H}_{r,k}\right){\mathbf{G}}.
\end{aligned}
\end{equation}

In this paper, we deploy a  uniform linear array (ULA) of antennas at the BS and a uniform planar array (UPA) of elements at the RIS, respectively. Considering the classical narrowband ray-based channel model in this paper, the \emph{user-independent} BS-RIS channel in spatial domain can be expressed as
\begin{equation}\label{eq6}
\mathbf{G}=\sum_{i=1}^{L_{1}} \alpha_{i} \mathbf{b}_{1}\left(\phi_{1,i},\theta_{1,i}\right)\mathbf{a}^{H}\left(\phi_{i}^{\mathrm{AoD}}\right),
\end{equation}
where ${L_{1}}$ is the number of dominant paths between the BS and the RIS,  $\alpha_{i}$ denotes the complex gain of the ${i}$-th path, and $\phi_{1,i}$ ($\theta_{1,i}$) denotes the azimuth  (elevation) angle-of-arrival (AoA) of the ${i}$-th path, respectively. Considering a UPA with $N_1$ horizontal elements and $N_2$ vertical elements ($N=N_1\times N_2$), the steering vector $\mathbf{b}_1\left(\phi_{1, i}, \theta_{1, i}\right)\in{\mathbb{C}}^{N \times 1}$ of the ${i}$-th path can be expressed as
\begin{equation}\label{eq8}
\begin{aligned} & \mathbf{b}_1\left(\phi_{1,i},\theta_{1,i}\right)=[\left.1, e^{j 2 \pi \frac{d_\mathrm{R}}{\lambda} \sin \theta_{1,i}}, \cdots, e^{j 2 \pi \frac{d_\mathrm{R}}{\lambda}\left(N_{2}-1\right) \sin \theta_{1,i}}\right]^{\mathrm{T}}   \otimes \\ & \frac{1}{\sqrt{N}}\left[1, e^{j 2 \pi \frac{d_\mathrm{R}}{\lambda} \cos \theta_{1,i} \sin \phi_{1,i}}, \cdots, e^{j 2 \pi \frac{d_\mathrm{R}}{\lambda}\left(N_{1}-1\right) \cos \theta_{1,i} \sin \phi_{1,i}}\right]^{\mathrm{T}},
\end{aligned}
\end{equation}
where $d_\mathrm{R}$ is the element spacing at the RIS, $\lambda$ is the wavelength of the carrier frequency.

The steering vector of the BS antenna array response  of the ${i}$-th path $\mathbf{a}\left(\phi_{i}^{\mathrm{AoD}}\right)\in\mathbb{C}^{M\times1}$ can be denoted as
\begin{equation}\label{eq9}
\begin{aligned} \mathbf{a}\left(\phi_{i}^{\mathrm{AoD}}\right)=\frac{1}{\sqrt{M}}\left[1, e^{j 2 \pi \frac{d_\mathrm{B}}{\lambda} \sin \phi_{i}^{\mathrm{AoD}}}, \cdots, e^{j 2 \pi \frac{d_\mathrm{B}}{\lambda}\left(M-1\right) \sin \phi_{i}^{\mathrm{AoD}}}\right]^{\mathrm{T}}, \end{aligned}
\end{equation}
where $d_\mathrm{B}$ denotes the antenna spacing at the BS, and $\phi_{i}^{\mathrm{AoD}}$ denotes the angle-of-departure (AoD) of the ${i}$-th path between the BS and the RIS.

Similar to (\ref{eq6}), the \emph{user-specific} RIS-UE channel ${\mathbf{h}}_{r,k}$ between the RIS and the ${k}$-th user  in spatial domain can be expressed as
\begin{equation}\label{eq5}
\mathbf{h}^H_{r, k}=\sum_{j=1}^{L_{2}} \beta_{k, j} \mathbf{b}_{2}^H\left(\phi_{2, k, j},\theta_{2, k, j}\right),
\end{equation}
where ${L_{2}}$ is the number of dominant paths between the RIS and the UE, $\beta_{k, j}$ is the complex gain of the ${j}$-th path, $\phi_{2, k, j}$ and $\theta_{2, k, j}$ are the azimuth and elevation AoDs of the ${j}$-th path, respectively, and $\mathbf{b}_{2}\left(\phi_{2, k, j},\theta_{2, k, j}\right)$ has the similar form with $\mathbf{b}_1\left(\phi_{1,i},\theta_{1,i}\right)$ in (\ref{eq8}).

According to (\ref{eq3}), (\ref{eq6}) and (\ref{eq5}), we denote the BS-RIS-UE cascaded channel~\cite{NadeemSystemModel}   of the $k$-th user in spatial domain as $\mathbf{H}_{k} \triangleq \mathrm{diag}\left({\mathbf{h}}^{H}_{r,k}\right){\mathbf{G}}$, which can be expressed as
\begin{equation}\label{eq7}
\begin{aligned}
\mathbf{H}_{k} = \sum_{i=1}^{L_{1}} \sum_{j=1}^{L_{2}}\alpha_{i}\beta_{k, j }   {\mathrm{diag}}  \left(\mathbf{b}_{2}^H \left(\phi_{2, k, j}, \theta_{2, k, j}\right)\right) & \\   \mathbf{b}_{1}\left(\phi_{1,i},\theta_{1,i}\right)&\mathbf{a}^{H}\left(\phi_{i}^{\mathrm{AoD}}\right).
\end{aligned}
\end{equation}

For simplicity, we rewrite (\ref{eq7}) as
\begin{equation}\label{eq11}
\mathbf{H}_{k} = \sum_{i=1}^{L_{1}} \sum_{j=1}^{L_{2}}g_{i, k, j} \mathbf{b}\left(\phi_{i, k, j}^{\mathrm{AoA}},\theta_{i, k, j}^{\mathrm{AoA}}\right) \mathbf{a}^{H}\left(\phi_{i}^{\mathrm{AoD}}\right),
\end{equation}
where $g_{i,k,j}\triangleq\alpha_{i}\beta_{k, j }$.

From (\ref{eq8}) and (\ref{eq7}), we can obtain the cascaded steering vector $\mathbf{b}\left(\phi_{i, k, j}^{\mathrm{AoA}},\theta_{i, k, j}^{\mathrm{AoA}}\right)\triangleq{\mathrm{diag}}\left(\mathbf{b}_{2}^H\left(\phi_{2, k, j},\theta_{2, k, j}\right)\right) \mathbf{b}_{1}\left(\phi_{1,i},\theta_{1,i}\right)$,  where $\phi_{i, k, j}^{\mathrm{AoA}}$  denotes the cascaded azimuth AoA at the RIS, whose sine value is the difference between the sine value of $\phi_{1,i}$ and $\phi_{2, k, j}$, and $\theta_{i, k, j}^{\mathrm{AoA}}$ denotes the cascaded  elevation AoA, whose sine value is the difference between the sine value of $\theta_{1,i}$ and $\theta_{2, k, j}$. And $(\phi_{i, k, j}^{\mathrm{AoA}},\theta_{i, k, j}^{\mathrm{AoA}})$ is termed as a pair of cascaded AoA.

\section{Proposed Dimension Reduced Channel Feedback Scheme}\label{S3}
In this section,  the single-structured sparsity of the BS-RIS-UE cascaded channel is  introduced at first. Then, a dimension reduced channel feedback scheme is proposed based on the single-structured sparsity.


\subsection{Single-Structured Sparsity of the BS-RIS-UE Cascaded Channel}\label{S3.1}
\textbf{Sparsity for a single user's BS-RIS-UE cascaded channel: }For RIS-assisted wireless communication system, the BS and the RIS are usually surrounded by limited scatterers. In other words, there are only a few AoDs at the BS and a few cascaded AoAs at the RIS. To illustrate this sparsity, we can rewrite the spatial domain channel (\ref{eq11})  as~\cite{LiangCE}
\begin{equation}\label{eq12}
\mathbf{H}_{k}=\mathbf{\tilde{H}}_{k} \mathbf{\Theta}_{T}^{H}=\mathbf{\Theta}_{R} \mathbf{A}_{k} \mathbf{\Theta}_{T}^{H},
\end{equation}
where $\mathbf{\Theta}_{R}\in{\mathbb{C}}^{N \times G_r}$ and $\mathbf{\Theta}_{T}\in{\mathbb{C}}^{M \times G_t}$ are the dictionary
matrices for the angular-domain channel with angular resolutions $G_r$ of the cascaded AoA at the RIS and $G_t$ of the AoD at the BS. Then, we can divide the cascaded AoA and AoD into $G_r$  and $G_t$ girds, respectively.

\begin{figure}[tp]
\begin{center}
\vspace*{0mm}\includegraphics[width=0.7\linewidth]{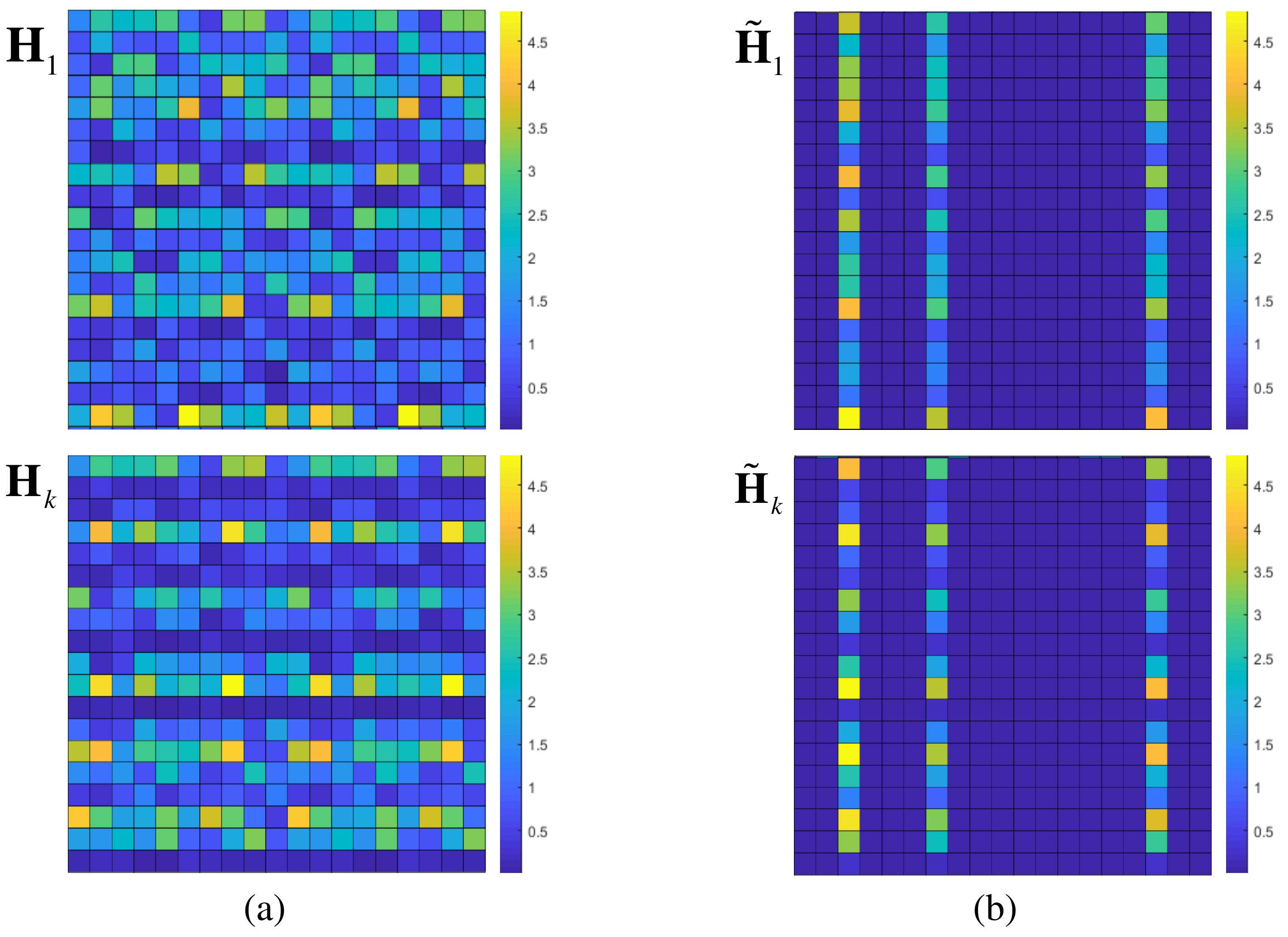}
\end{center}
\vspace*{-3mm}\caption{The BS-RIS-UE cascaded channels (a) in the spatial domain and (b) in the hybrid domain.} \label{FIG2}
\vspace*{-2mm}
\end{figure}

$\mathbf{H}_{k}$ and $\mathbf{A}_{k}$ are the spatial-domain  and angle-domain channel respectively, which are familiar in the existing research. In this paper, we denote $\mathbf{\tilde{H}}_{k}$  as the hybrid domain (hybrid spatial- and angular-domain) channel, since the row dimension of $\mathbf{\tilde{H}}_{k}$ is converted into the angular domain by $\mathbf{\Theta}_{T}$, while the column dimension is still in the spatial domain. As can be seen from Fig. \ref{FIG2}, there are only  $L_1$ non-zero columns in hybrid domain channel $\mathbf{\tilde{H}}_{k}$, since
there are  $L_1$ AoDs at the BS. Besides, the indexes of the non-zero column positions are corresponding to the gird indexes of the AoDs.

To reduce the overhead for channel feedback,  we can  focus on feeding the non-zero columns of $\mathbf{\tilde{H}}_{k}$ back to the BS. According to (\ref{eq11}) and (\ref{eq12}), the $i$-th   non-zero of $\mathbf{\tilde{H}}_{k}$, which is also the $g_t$-th column of $\mathbf{\tilde{H}}_{k}$ can be esxpressed as
\begin{equation}\label{eq13}
\mathbf{\tilde{h}}_{k,i}\triangleq\mathbf{\tilde{H}}_{k,(:,g_t)} =  \sum_{j=1}^{L_{2}}g_{i, k, j} \mathbf{b}\left(\phi_{i, k, j}^{\mathrm{AoA}},\theta_{i, k, j}^{\mathrm{AoA}}\right) ,
\end{equation}
where $g_t=1,\cdots,G_t$ and $i=1,\cdots,L_1$. Besides, (\ref{eq13}) can be rewrited as $\mathbf{\tilde{h}}_{k,i}= \mathbf{B}_{k,i}\mathbf{g}_{k,i}$,
where $\mathbf{B}_{k,i}=\left[\mathbf{b}\left(\phi_{i, k, 1}^{\mathrm{AoA}},\theta_{i, k, 1}^{\mathrm{AoA}}\right), \cdots, \mathbf{b}\left(\phi_{i, k, L_2}^{\mathrm{AoA}},\theta_{i, k, L_2}^{\mathrm{AoA}}\right)\right]\in{\mathbb{C}}^{N \times L_2}$ and $\mathbf{g}_{k,i}=\left[g_{i, k, 1}, \cdots, g_{i, k, L_2}\right]^T\in{\mathbb{C}}^{L_2 \times 1}$. There is a brief explanation for (\ref{eq13}): Most of columns in $\mathbf{\Theta}_{T}$ are orthogonal to the  steering vector $\mathbf{a}\left(\phi_{i}^{\mathrm{AoD}}\right)$, except when the angle gird of  column vector is equal to the angle of steering vector.


\textbf{Single-structured sparsity of different users' BS-RIS-UE cascaded channels: }The BS-RIS-UE cascaded channel consists of two parts: the \emph{user-independent} BS-RIS channel and \emph{user-specific} RIS-UE channel, i.e., different users share the same channel between the BS and RIS, but differ in channels between the RIS and UEs. In other words, different UEs have the same   AoDs at the BS, but  unique  cascaded AoAs at the RIS.  To sum up, for the BS-RIS-UE cascaded channels of different users, the sparsity is structured only in the column dimension but not in the row dimension, which is termed as single-structured sparsity in this paper\footnote{This channel characteristic has also been used to improve the channel estimation accuracy in~\cite{LiangCE}.}.

As shown in Fig. \ref{FIG2}(b), the number of non-zero columns are limited, and different UEs' non-zero columns     have the same indexes, i.e., the sparsity is structured for different UEs in the column dimension, since we transform the user-independent AoDs with dictionary matrix. However, if we  transform the user-specific AoAs  with another dictionary matrix, we will acquire a few non-zero rows, but the indexes of non-zero rows differ from different UEs, i.e., the  structured sparsity appears only in the single side.



\subsection{Proposed Dimension Reduced Channel Feedback Scheme}\label{S3.2}
In this subsection, we propose a dimension reduced  channel feedback scheme by utilizing the single-structured sparsity. We only focus on the BS-RIS-UE cascaded  channel  feedback   in this paper\footnote{Because there is no difference between the direct BS-UE channel in RIS-assisted scneario and the conventional wireless channel in existing scneario.}.  Besides, we can just feed the non-zero column vectors  hybrid channel $\mathbf{\tilde{H}}_{k}$     by turns with vector codebooks to reduce the overhead. In the proposed scheme,  the codebook is designed by cascaded AoAs, which will be introduced in step 2 later.

\begin{figure}[tp]
\begin{center}
\vspace*{0mm}\includegraphics[width=1\linewidth]{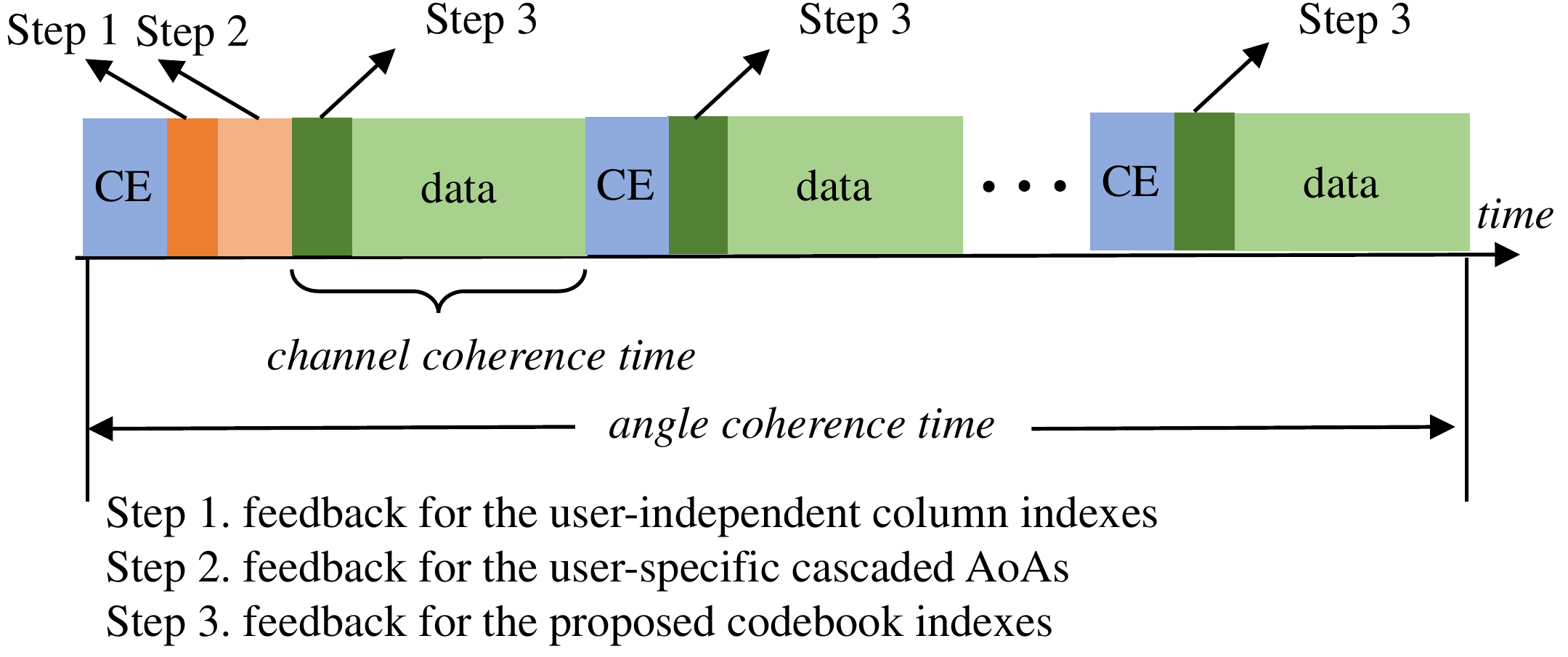}
\end{center}
\vspace*{-3mm}\caption{The proposed dimension reduced channel feedback frame structure. The downlink CSI should be acquired by channel estimation at the UE firstly. Like other channel feedback research, we assume that the UE can acquire  downlink CSI perfectly~\cite{FBSubspace1(ShenTcom)}.} \label{FIG3}
\vspace*{-3mm}
\end{figure}

Hence,  what we need to feed back are the position indexes of non-zero columns, the cascaded AoAs and the indexes of codeword, as shown in Fig. \ref{FIG3}.
Specifically, $\mathbf{\tilde{H}}_{k}\left(k=1,2,\cdots,K\right)$ can be fed back to the BS by three steps in our proposed scheme. At first, we feed the indexes of non-zero columns in $\mathbf{\tilde{H}}_{k}$ back to the BS.    Then, the user-specific cascaded AoAs, which can be used to design the vector codebook, need to be quantized and fed back to the BS. Finally, we select the appropriate codeword from the designed codebook for every non-zero column vector of $\mathbf{\tilde{H}}_{k}$, and feed the corresponding codeword index back to the BS.  After receiving  these parameters, we can easily construct the feedback channel matrix  at the BS.


To be more specific, we describe the three steps as follows:


\textbf{Step 1. Feedback for the user-independent indexes of non-zero columns.} In this step,  the position indexes of the non-zero columns of the BS-RIS-UE cascaded channel $\mathbf{\tilde{H}}_{k}$ will be acquired and fed back to the BS. According to (\ref{eq12}), the angular resolution of the dictionary matrix $\mathbf{\Theta}_{T}$ is  $G_t$, i.e., the AoDs are quantified by $G_t$ girds. Hence, the index of the AoD quantized gird is also the index of the non-zero column of $\mathbf{\tilde{H}}_{k}$. As shown in Fig. \ref{FIG3},  AoDs can be acquired by channel estimation at first. Then, the  gird angle index is fed back to the BS. Thanks to the single-structured sparsity,  we just need to appoint one user  to feed the ${L_{1}}$ user-independent column indexes back to the BS, while other users can skip this procedure. In this way, the channel feedback overhead can be significantly  reduced  from the following two aspects: 1) We can reduce the number of users who participate the index feedback  from $K$  to only one by utilizing the single-structured sparsity.  2) For this user, by utilizing the property of hybrid domain channel, the number of indexes which need to be fed back, can be reduced from the number of antennas $M$ to  the dominant paths $L_1$, which is usually much smaller than the former one.

\textbf{Step 2. Feedback for the user-specific cascaded AoAs.} To feed back the hybrid domain channel $\mathbf{\tilde{H}}_{k}$, we divide  $\mathbf{\tilde{H}}_{k}$ to several  non-zero columns. Although the non-zero column of $\mathbf{\tilde{H}}_{k}$ is not sparse, we can still utilize the angle-adaptive subspace~\cite{FBSubspace1(ShenTcom)} to design the codebook with a low overhead. In this step, the cascaded AoAs which acquired by  channel estimation,  will be quantized as $B_0=7$ bits and fed back to the BS one by one. The $i$-th non-zero column of   the $k$-th user $\mathbf{\tilde{h}}_{k,i}$  corresponds to $L_2$ pair cascaded AoAs. According  to the $L_2$ pair quantized cascaded AoAs $\{(\hat{\phi}_{i, k, j}^{\mathrm{AoA}},\hat{\theta}_{i, k, j}^{\mathrm{AoA}})\}_{j=1}^{L_2}$,  we can design a codebook for this  non-zero column.  As shown in (\ref{eq13}), the $i$-th non-zero column of $\mathbf{\tilde{H}}_{k}$ is a $N$-dimensional vector, which can be composed by $L_2$ (usually smaller than $N$) paths between the RIS and the $k$-th user. In other words,  $\mathbf{\tilde{h}}_{k,i}$ is distributed in the $L_2$-dimensional subspace of the $N$-dimensional channel space. By utilizing this characteristic, we generate the codeword  with a steering matrix and low-dimensional codeword vector from random vector quantization (RVQ)~\cite{FBCodebook1(RVQ)} codebook.

The $q$-th ($q=1,\cdots,2^B$) codeword of the generated codebook $\mathcal{C}$   can be expressed as
\begin{equation}\label{eq15}
\mathbf{c}_{k,i,q}= \hat{\mathbf{B}}_{k,i}\mathbf{r}_{k,i,q},
\end{equation}
where $\mathbf{r}_{k,i,q}\in\mathbb{C}^{L_2\times1}$ is a low-dimensional codeword vector chosen from the conventional RVQ codebook with size as $2^B$, and $\hat{\mathbf{B}}_{k,i}\in{\mathbb{C}}^{N \times L_2}$ is the steering matrix which can be expressed as
\begin{equation}\label{eq19}
\hat{\mathbf{B}}_{k,i}=\left[\mathbf{b}\left(\hat{\phi}_{i, k, 1}^{\mathrm{AoA}},\hat{\theta}_{i, k, 1}^{\mathrm{AoA}}\right), \cdots, \mathbf{b}\left(\hat{\phi}_{i, k, L_2}^{\mathrm{AoA}},\hat{\theta}_{i, k, L_2}^{\mathrm{AoA}}\right)\right].
\end{equation}

\textbf{Step 3. Feedback for the generated codebook indexes.} In this step, we focus on feeding back the  codeword indexes  for the ${L_{1}}$ non-zero columns of $\mathbf{\tilde{H}}_{k}$, by utilizing the non-zero column indexes fed back in the step 1 and  the  subspace codebooks generated in the step 2. Specifically, taking $\mathbf{\tilde{h}}_{k,i}$ as an example, from the  corresponding subspace codebook, we select the codeword for $\mathbf{\tilde{h}}_{k,i}$ which satisfies  the constraint as $D=\underset{q \in\left[1,2^{B}\right]}{\arg \min } \sin ^{2}\left(\measuredangle\left(\mathbf{\tilde{h}}_{k,i}, \mathbf{c}_{k,i,q}\right)\right)$ at first. Then the  index $D$ of this codeword needs to be fed back to the BS with $B$ bits. In this paper, we adopt  the value of $B$ by utilizing the simulated analysis. Similarly, the codeword indexes corresponding to other non-zero columns can be fed back  in the same way.

According to the feedback parameters of the above three steps, the BS can recover the BS-RIS-UE cascaded channel $\mathbf{\tilde{H}}_{k}$ as follows. The BS can recover the non-zero column vector $\mathbf{\tilde{h}}_{k,i}$ according to the vector codebook from step 2 and the codeword index from step 3. After recovering $L_1$ non-zero columns, $\mathbf{\tilde{H}}_{k}$ will be obtained by arranging the column vectors according to the position  indexes from step 1. Finally, the spatial domain  $\mathbf{H}_{k}$ can be acquired with the help of (\ref{eq12}).

Note that the  scatterers around the BS and the RIS remain unchanged over a long time scale, which is termed as the angle coherence time~\cite{FBSubspace1(ShenTcom)}, in which the angles remain unchanged, is larger than the channel coherence time.  Hence, during the angle coherence time,   step 1 and step 2  only need be performed once, which reduce the overhead further, as shown in Fig. \ref{FIG3}.

\section{Simulation Results}\label{S4}

For the simulation setup,   the number of BS antennas, RIS  elements, and single-antenna users are $M=32$,  $N=64$, and $K=4$, respectively. The number of paths are set as $L_1=4$ and $L_2=2$, respectively. The  SNR at receiver is set to 5 dB.

After acquiring the  downlink CSI via channel feedback, cross entropy optimization (CEO)~\cite{GaoCEO} is utilized to optimize the joint beamforming at the BS and the RIS. We choose the per-user rate to evaluate the performance of the channel feedback scheme, which can be expressed as $R=\mathrm{E}\left[\log _{2}\left(1+\frac{\frac{\gamma}{K}\left|{\mathbf{h}}^{H}_{\mathrm{DL},k} \hat{\mathbf{v}}_{k}\right|^{2}}{1+\frac{\gamma}{K} \sum_{i=1, i \neq k}^{K}\left|{\mathbf{h}}^{H}_{\mathrm{DL},k} \hat{\mathbf{v}}_{i}\right|^{2}}\right)\right]$,
where $\gamma$ is the transmit power; $\hat{\mathbf{v}}_{k} \in \mathbb{C}^{M\times1}$ is the $k$-th column of the normalized precoding matrix at the BS. And the phase-shift diagonal matrix ${\mathbf{\Phi}}$ for the RIS has been included in $\mathbf{h}^{H}_{\mathrm{DL},k}$, as shown in (\ref{eq3}). The per-user overhead is the total feedback overhead divided by number of UEs.

\begin{figure}[tp]
\begin{center}
\vspace*{-3mm}\includegraphics[width=0.9\linewidth]{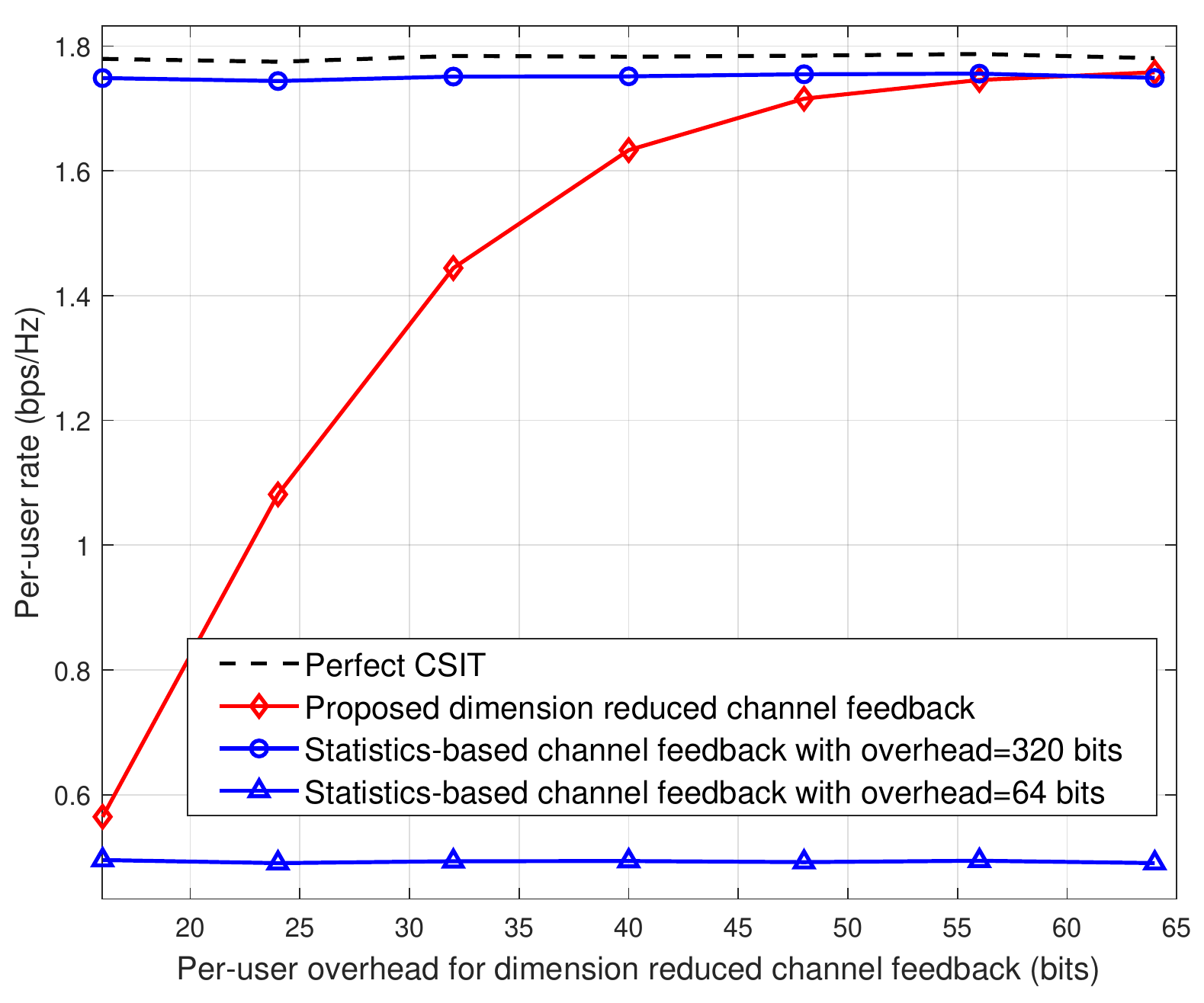}
\end{center}
\vspace*{-3mm}\caption{Comparison of the per-user rate between the dimension reduced channel feedback scheme and the conventional feedback scheme.} \label{FIG4}
\vspace*{-3mm}
\end{figure}
Fig. \ref{FIG4} shows the per-user rate  versus the per-user feedback overhead. We compare our performance with a conventional channel statistics-based feedback scheme in~\cite{FBSubspace1(ShenTcom)}. And we also give the upper bound  with the perfect channel state information at the transmitter (CSIT). In this figure, we choose the AoD resolution $G_t=512$ and the quantized bits of the cascaded AoA as $B_0=7$ bits. The per-user overhead rises with the size of codeword index $B$ increasing from 1 to 13 bits. With the per-user overhead equal to  64 bits, the proposed scheme can acquire the near-optimal performance. However, to achieve the same performance, the per-user overhead of the conventional scheme is 320 bits, which means the proposed scheme can reduce the overhead about 80\% while the performance can still be guaranteed. Besides, with 64 bits in the conventional scheme, the per-user rate is only 28\% compared with the proposed scheme.

\begin{figure}[tp]
\begin{center}
\vspace*{-3mm}\includegraphics[width=0.9\linewidth]{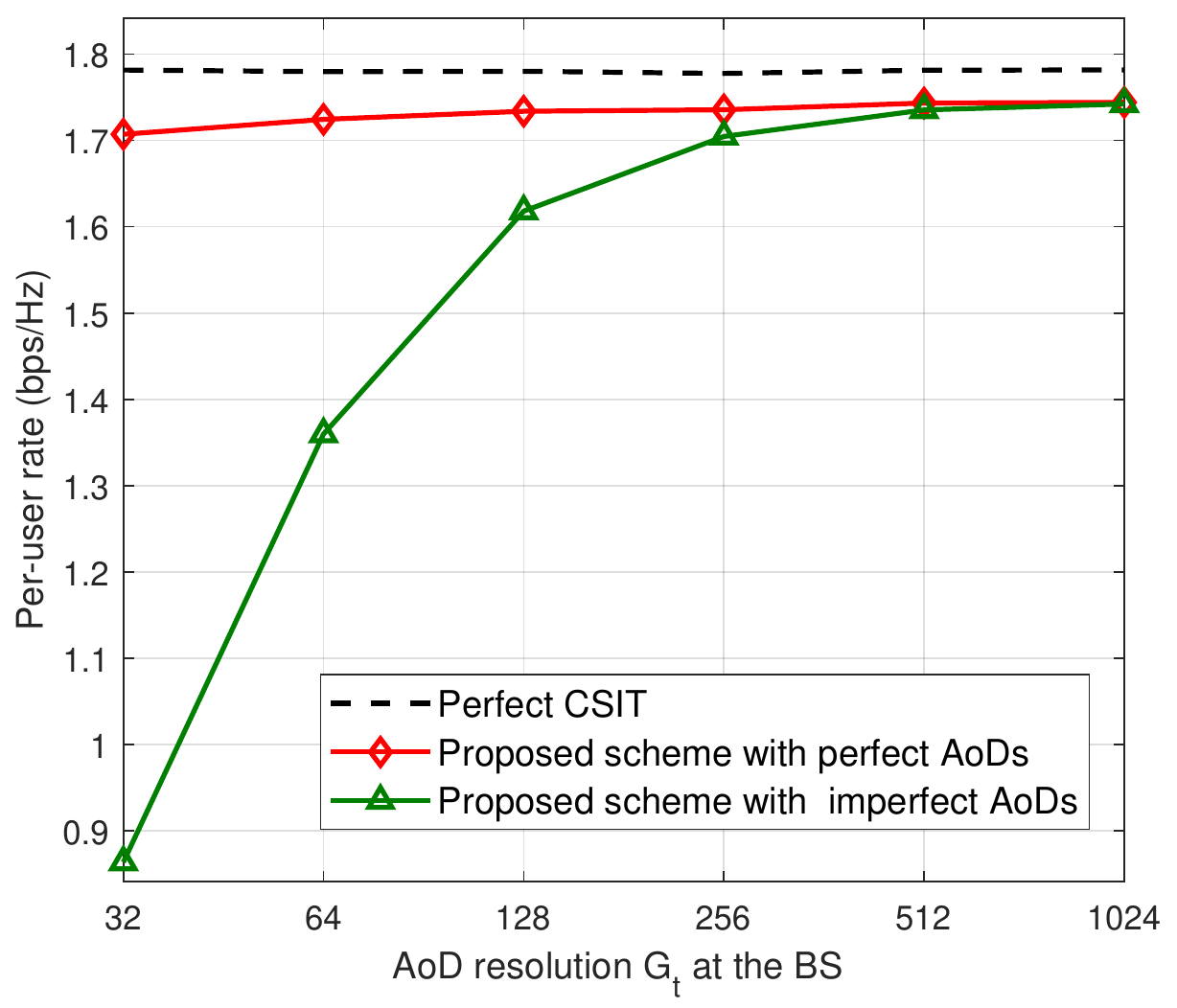}
\end{center}
\vspace*{-3mm}\caption{The per-user rate of the dimension reduced channel feedback scheme against the   AoD resolution G$_t$ at the BS.} \label{FIG5}
\vspace*{-3mm}
\end{figure}

Fig. \ref{FIG5} illustrates the impacts of the AoD resolution $G_t$ with $B=10$ bits and $B_0=6$  bits. To feed the single-structured indexes back to the BS in step 1, AoD at the BS must be approximated to the nearest gird.  From Fig. \ref{FIG5} we can find that when $G_t=512$, the performance with imperfect AoDs on gird can almost attain the per-user rate of the proposed scheme with perfect AoDs. Meanwhile, considering the angle coherence time, the corresponding overhead for channel feedback can be reduced further. For example, thanks to the single-structured sparsity, even if 25 \% of users are selected to feed the AoD indexes back to the BS to improve the  robustness of channel feedback (although we only need to select one user theoretically speaking), the per-user feedback overhead of step 1 is $L_1\times\mathrm{log}_2(G_t)\times 25\%/10\approx1$ bit, when the angle coherence time is 10 times of the channel coherence times.

\section{Conclusions}\label{S5}
In this paper, we investigate the channel feedback problem in RIS-assisted wireless communications for the first time to resist the unbearable channel feedback overhead in the RIS-assisted wireless communications. Firstly, we exploit the single-structured sparsity of the BS-RIS-UE cascaded channel. Then, by utilizing this channel characteristic, a dimension reduced channel feedback scheme consisting of three steps  are proposed for the BS-RIS-UE cascaded channel feedback.  Besides, by leveraging the angle coherence time, the overhead can be reduced further. Simulation results show that the dimension reduced channel feedback scheme can reduce the channel feedback overhead dramatically, while the near-optimal per-user rate  can be guaranteed. We leave the delicate codebook design by exploiting the single-structured sparsity for future works.
\section*{Acknowledgment}
This work was supported by the National Science and Technology Major Project of China under Grant 2018ZX03001004-003 and the National Natural Science Foundation of China for Outstanding Young Scholars under Grant 61722109.

\footnotesize
\balance
\bibliographystyle{IEEEtran}
\bibliography{IEEEabrv,Shen1Ref}
\end{document}